\documentclass[twocolumn,amsmath,amssymb,10pt,a4paper]{revtex4}

\usepackage{graphicx}
\usepackage{dcolumn}
\usepackage{bm}
\begin{document}
\title{Statistics of collective human behaviors observed in blog entries}
\author{Yukie Sano$^1$}
\author{Kimmo Kaski$^2$}
\author{Misako Takayasu$^1$}

\affiliation{$^1$Department of  Computational  Intelligence  and  Systems  Science,
Interdisciplinary Graduate School of Science and Engineering,
Tokyo Institute of Technology, 4259-G3-52 Nagatsuta-cho, Midori-ku, Yokohama 226-8502, Japan}
\affiliation{$^2$Centre of Excellence in Computational Complex Systems Research, 
Department of Biomedical Engineering and Computational Science, 
Helsinki University of Technology, P.O.Box 2200, FI-02015 TKK, Finland }


\maketitle
\section{Introduction}
\label{intro}
A new field attracts the attention of physics researchers 
when precise quantitative observation becomes available. 
A good example of this is financial market price fluctuation 
that was not among physicists' interests twenty years ago.
However, the huge number of precise observations 
and numerical simulation techniques 
led to a new field of applied statistical physics by the name of econophysics~\cite{eco}.
More recently, the sales figures for popular books have been analyzed 
in view of the universal responses of complex systems~\cite{sornette}. 
Further, the increase in the number of registrations for a conference as 
the deadline nears has been studied and concluded to a result of the 
superposition of random processes caused by naive human behaviors~\cite{alfi}.

At an individual level, any human behavior may be recognized as intentional and  
not random. However, by observing a large number of people simultaneously, 
we can expect to observe the random nature of the gross properties owing to mutual independence. 
The first step in the data analysis of human behaviors is to clarify 
how real human system shows randomness by comparing it with predictions, 
based on the assumption of independence. 
The next step is to establish an empirical relation 
about the gap between the simple theory and observation. 
The third step is to develop a new theory that captures a deeper human action or hidden social interaction.

 For example, paying attention to the time intervals of deals in financial markets, 
an exponential distribution of intervals is naturally expected as the occurrence of a deal 
is caused by a collision of buy and sell orders at the same price, 
which can be modeled by a Poisson process. 
However, in reality we generally observe a time interval 
distribution characterized by a fat-tail. 
It is clarified from data analysis that the underlying Poisson parameter is changing 
in the time scale of a few minutes and the observed fat-tailed distribution is given 
by the superposition of exponential distributions 
with various characteristic scales~\cite{takayasu1}. 
The human behavior underlying this phenomenon is considered to be 
a general tendency of market participants to modulate their 
own clock speed to be proportional to a moving average of the market's transaction intervals. 
This effect, named self-modulation, is known to induce long-term correlation and 
make the power spectrum of events follow a $1/f$ spectrum~\cite{takayasu2}. 
An agent-based numerical model can reproduce this effect and we can find that the resulting 
transaction intervals are compatible with real data~\cite{yamada}.

\begin{figure}
\includegraphics{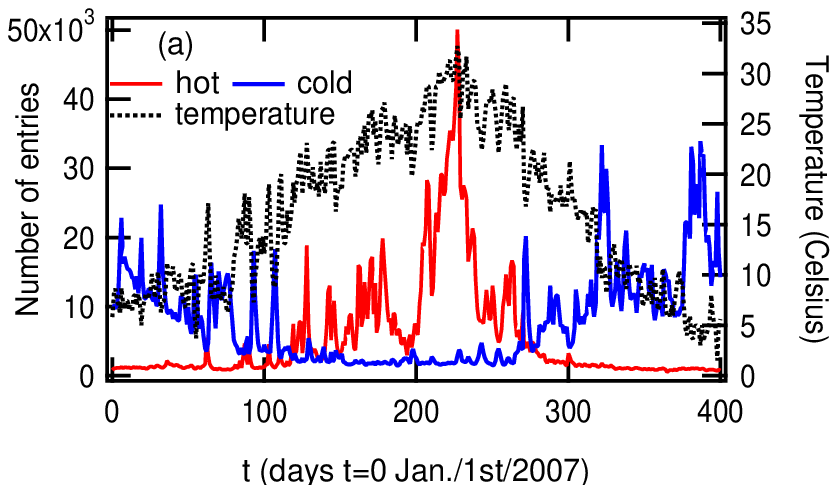}
\includegraphics{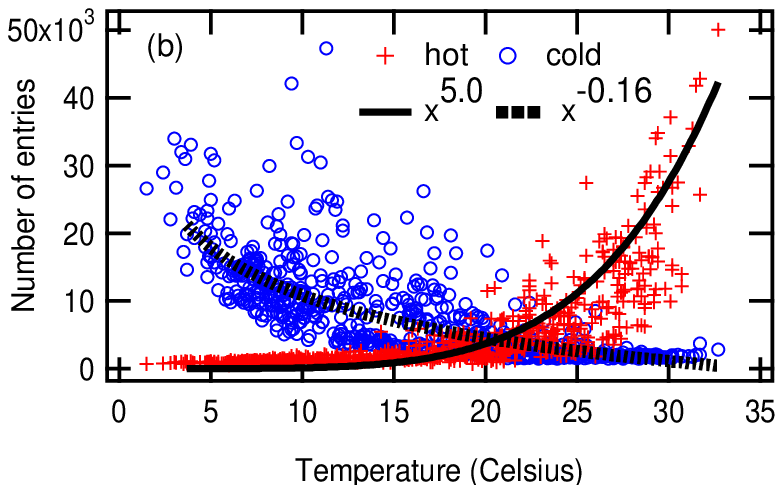}
\caption{\label{fig:Fig0}Relation between temperature and number of blog entries.
(a) Time series of tempetature and number of blogs including word ``hot" and ``cold". 
(b) Scatter plot of temperature and number of blogs.}
\end{figure}

 In this paper, we focus on the data analysis of blogs to observe collective human behaviors. 
A blog is a new type of social communication medium where personal impression can be easily uploaded on the web. 
A blog first appeared in the late 1990s and rapidly gained popularity. 
The number of blogs has been increased every year since then. 

 A typical blog is maintained by an individual or a small group. 
It consists of entries posted by users that contain text and/or images 
and are frequently accompanied by links to other web pages. 
One of the characteristics of a blog is that 
readers can easily leave their comments in an interactive format. 
Consequently, a blog has evolved as a new type of social communication tool for individuals.

 Search engine technologies have been developed 
to observe the details of blog entries automatically at high speeds~\cite{kizasi,google}. 
We focus on the temporal changes in the frequency of a word on the web. 
For a given keyword, the search engine automatically lists all blog entries 
that include the keyword along with the time of posting. 
According to Technorati~\cite{technorati}, 
an internet search engine company for blogs, 
the number of blogs in the world is now more than 70 million and 
this figure is still increasing. Categorizing blogs in terms of language, 
we find that in 2007, 37\% of the blogs were in Japanese, 
36\% in English, 8\% in Chinese, etc.~\cite{report}.
Here, we observe Japanese blogs from January 1st, 2007 to December 31st, 2008 
using a search engine provided by Dentsu Buzz Research~\cite{dentsu}. 
This search engine covers 20 major blog providers in Japan 
covering more than 10 million users. 
The total number of observed entries exceeds 600 million; 
in other words, on average, about 800 thousand new entries are uploaded every day.

Here, we show an example that blogs are often reflect peoples reaction to the 
social and natural phenomena. In Fig.\ref{fig:Fig0}, we show relationship between the average daily temperature in Tokyo and 
number of blog entries including the keywords ``hot" and  ``cold". 
In summer time, in the middle of Fig.\ref{fig:Fig0}(a), it is getting hot and 
people tend to post blog entries with keyword ``hot" more frequently. 
We confirm that there is a nonlinear relation between real temperature and number of entries by the scatter plot [Fig.\ref{fig:Fig0}(b)].

\section{Basic statistics}
\label{sec2}
A challenging blog analysis is applied for early detection of epidemics. 
Ginsberg \textit{et al}. observed keywords such as ``flu" by using the search engine query data 
on the Google homepage and confirmed a positive correlation between the actual spread 
and word appearances~\cite{ginsberg}. Here, we focus on the more basic properties of word frequency.

A pioneering basic study of blog word appearance has been done by Lambiotte \textit{et al}. 
who analyzed appearance intervals of common words~\cite{lambiotte}.
It was reported that a naive Poissonian assumption fails in general 
even for low-frequency words. 
However, two of the authors (Y.S. and M.T.) recently examined 
similar time series of blog word appearances carefully 
and found that such deviation from the Poisson process is caused by the following three effects: 
\begin{enumerate}
\item Repeated spamming. 
\item Non-stationary changes caused by server breakdowns or the growth in blog population. 
\item Periodic fluctuations during specific days of the week.
\end{enumerate}
By introducing normalization processes to eliminate these effects, 
we can confirm that, as expected theoretically, low-frequency words actually follow the 
Poisson process~\cite{sano}. 
It is interesting that we still observe deviation from the Poisson process for any high-frequency words. 
We focus on such high-frequency words. 

\begin{figure}
\includegraphics{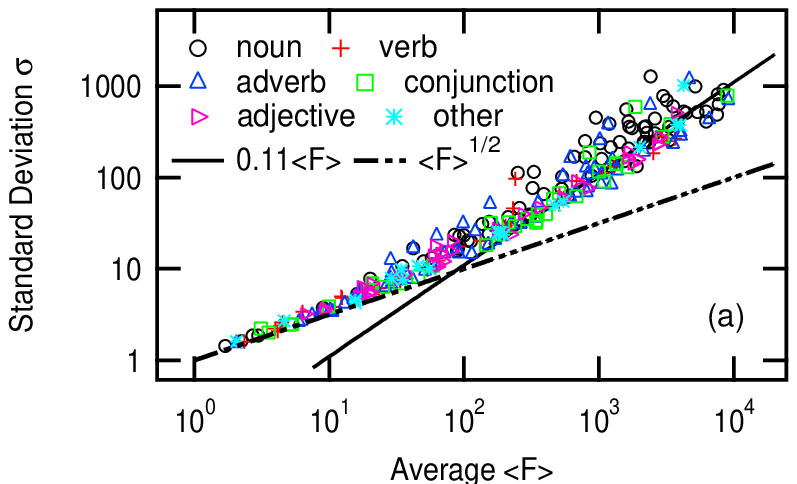}
\includegraphics{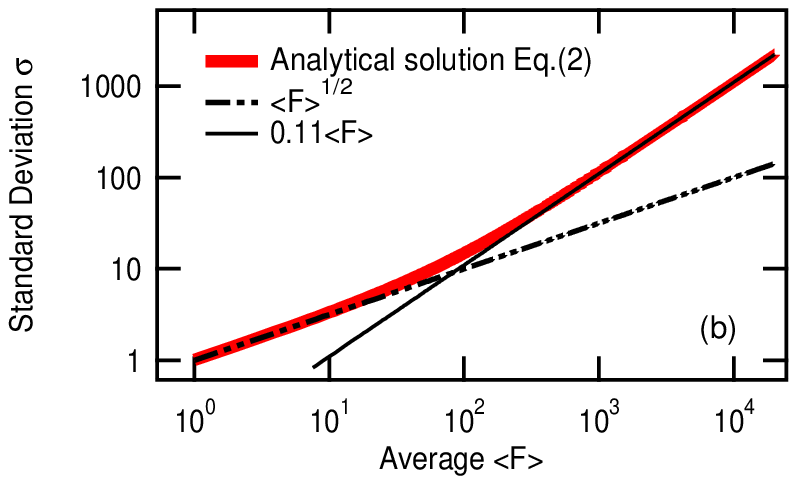}
\caption{\label{fig:Fig1}Relation between average $\langle F \rangle$ and standard deviation 
$\sigma$ in log-log scale. The dashed line corresponds to 
$\sigma=\sqrt{\langle F \rangle}$ and the solid line, to 
$\sigma=0.11\langle F \rangle$. 
(a) Empirical result for the 271 words. 
(b) Comparison with analytical solution of Eq.(\ref{eq:2}) (bold solid line).}
\end{figure}

In order to clarify the reason for this deviation from the Poisson process, 
we introduce the following data analysis. 
We randomly chose 300 words from a dictionary of Japanese morphological analysis. 
We get rid of 29 words from these words in case that the time series has 5 times higher peak from the average or 
average is more than $10^4$. Then, we get 78 nouns, 36 verbs, 66 adverbs, 37 adjectives, 
35 conjunctions, and 19 other types of words. 
For the $j$-th word, we observe the daily number of blogs that contain the $j$-th word 
by using the search engine $x_j(t)$. Then, we apply the above-mentioned normalization procedures 
to get a normalized blog number time series $F_j(t)$.
The average over the entire observation time $\langle F_j \rangle$ 
and the standard deviation $\sigma_j=\sqrt{\langle \left(F_j - \langle F_j \rangle\right)^2\rangle}$ 
is plotted for each word in a log-log scale in Fig.\ref{fig:Fig1}(a).

 In the case that the appearance of a word is approximated by the Poisson process, 
the standard deviation is given by the square root of the average value, 
and a line with slope 1/2 is expected. 
For words with frequencies less than about 80 entries per day, 
this relation can be confirmed to be independent of the morphological category 
from the left part of Fig.\ref{fig:Fig1}(a).
For these words, the autocorrelations of intervals are confirmed to be almost 0 
and the frequency distributions are checked to pass statistical tests of the Poisson distribution; 
namely, the appearance of low-frequency words can be approximated by the Poisson process. 
It should be noted that this result is so delicate that 
it cannot be obtained without the above normalization procedure, 
which might have been missing in the preceding study~\cite{lambiotte}.

For more-frequently appearing words, we can find a clear deviation from the simple Poisson process. 
As shown in the right part of Fig.\ref{fig:Fig1}(a), instead of a square root relation, 
a linear relation between standard deviation 
and average holds empirically.
Similar linear relations between standard deviation and average 
are observed in various fields such as river flow fluctuations, 
traffic fluctuations on highways~\cite{menezes} and numerical fluctuations in cellular molecules~\cite{kaneko}. 
In the case of traffic fluctuations, this relation 
is understood by a simple model named Random Diffusion (RD) model~\cite{meloni}, 
where random walkers are randomly injected into a given network. 
By observing the flow of walkers at a node, the linear relation between 
average flow strength and standard deviation is confirmed for large value of flow. 
Here, we introduce a simple model of posting blog entries as an application. 

We assume the situation that there is no spam and the system is stationary with no weekly period. 
There are $n(t)$ bloggers who are actively posting blogs at time $t$. 
Each blogger independently posts a blog entry containing the $j$-th keyword with probability $c_j$. 
The number of bloggers $n(t)$ is assumed to change randomly in the range of $[N-\Delta, N+\Delta]$, 
where $\Delta$ is a positive constant. 
Thus, the mean number of bloggers who post a blog entry containing the $j$-th word at time $t$ is $F_j(t)=c_jn(t)$.
Considering the case of a uniformly distributed $n(t)$ for simplicity, 
we obtain the probability density of the number of keyword appearances $p_j(F_j)$ 
by taking the superposition of the Poisson distribution.

\begin{equation}
p_j(F_j)=\sum_{k=0}^{2 \Delta}\frac{e^{-c_j (N- \Delta+k)}}{2 \Delta +1}\frac{[c_j(N- \Delta+k)]^{F_j}}{F_j!}.
\label{eq:1}
\end{equation}

From this probability density, the standard deviation $\sigma_j$ 
is expressed as a function of the average value $\langle F_j \rangle$, 
and the relation for large $\Delta$ is approximated as follows.

\begin{align}
\sigma_j &=\sqrt{\langle F_j \rangle \left(1+ \langle F_j \rangle \frac{\Delta (\Delta +1)}{3N^2}\right)} 
\notag \\
& \approx
\begin{cases}
  \sqrt{\langle F_j \rangle } & \langle F_j \rangle \ll \frac{3N^2}{\Delta^2} \\
  \frac{\Delta}{\sqrt{3}N}\langle F_j \rangle & \langle F_j \rangle \gg \frac{3N^2}{\Delta^2}
 \end{cases}.
\label{eq:2}
\end{align}

By fitting this theoretical estimation to the empirical relation in Fig.\ref{fig:Fig1}(a), 
we can tune the model's parameters. Since from Fig.\ref{fig:Fig1}(a), 
the bending point is estimated as ${\langle F\rangle}_c=82$, and from Eq.(\ref{eq:2}), 
it is given as $\langle F\rangle_c=3N^2/\Delta^2$, we have $\Delta/N=0.19$. 
The linear relation between standard deviation and average in Eq.(\ref{eq:2}) 
becomes $\sigma=0.11\langle F \rangle$, 
which automatically fulfills the empirical relation as shown in Fig.\ref{fig:Fig1}(b). 
This result demonstrates that the non-trivial relation 
between standard deviation and average is a general property 
caused by the daily fluctuations in the number of bloggers. 

 Although the above theoretical analysis captures the basic relation between the average number of keywords 
and the standard deviations, the data points in Fig.\ref{fig:Fig1}(a) scatter largely for words with 
large averages. It is confirmed that this scattering cannot be explained by the above superposition of 
stationary Poisson processes, and we need to consider the effect of temporal change of the appearance 
probability $c_j$.

\begin{figure}
\includegraphics{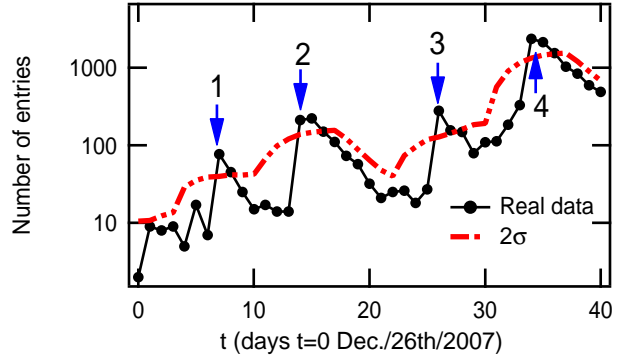}
\caption{\label{fig:Fig2}Example of automatic detection of peculiar fluctuations.}
\end{figure}

\section{Application}
\label{sec3}
As an application of this linear relation between 
standard deviation and average, here, we introduce a
method of finding extraordinary events hidden in fluctuations
of word appearances in blogs. We pay attention to the
name of new bread, which went on the market on
January 29th, 2008. The time series of the daily number of
blogs containing this new name after normalization
is given in Fig.\ref{fig:Fig2}. Before December 26th, 2007, the number
count of this keyword was 0. Later, the number count grew
rapidly as though in a type of oscillation. In order to
remove the weekly period, we introduce a 7-day moving average
which is defined by

\begin{equation}
\overline{F_j(t)}=\frac{1}{2M+1} \sum _{i=-M}^M{F_j(t+i)},
\label{eq:3}
\end{equation}
where $M=3$. Then, we estimate the standard deviation around this value 
based on the above-mentioned empirical relation as follows:

\begin{equation}
\sigma_j(t)=\sqrt{\overline{F_j(t)}\left(1+ \overline{F_j} \frac{\Delta^2}{3N^2}\right)},
\label{eq:4}
\end{equation}
where $\Delta/N=0.19$. The daily number count for this keyword is 
mostly within the $2\sigma(t)$ band. 
However, occasionally it is higher. 
We traced back the news and found that 
external factors such as new items released on those days did have an effect:

\begin{enumerate}
\item Jan./2nd/2008: One of the popular blogs introduced the name of this new bread.
\item Jan./9th/2008: The company put out a news release announcing the new bread.
\item Jan./21st/2008: The name of this new bread was introduced in a TV program.
\item Jan./29th/2008: The new bread went on sale.
\end{enumerate}

By using this empirical standard deviation, 
we can successfully detect the peculiar fluctuations that were apparently caused by the above external factors.
Furthermore, we can evaluate quantitatively the impact of the company's press releases to people.

\section{Discussion}
\label{sec4}
In summary, we first showed the importance of noise reduction for quantitative analysis of blog-word frequencies. 
For low-frequency words, we confirmed the Poisson hypothesis. 
For more-frequent words, we found that the effect of the change in the number of contributors 
modifies the distribution drastically 
and the standard deviation of blog-number fluctuations becomes proportional to 
the average values. 
Analytical solutions successfully 
reproduce the above basic properties of blog-number fluctuations. 

In this research we only observed Japanese blogs; however, 
we expect that the analysis method and the numerical model is directly 
applicable to other languages. 
By establishing the basic observation methods and simulation models, 
blog analysis will become a powerful tool for the scientific study of collective human behaviors in the society.
\\

\section*{Acknowledgements}
The authors appreciate the cooperation of Dentsu, Inc. and hottolink, Inc. for providing the blog data.

\end{document}